# Data Storage: Review of Heusler Compounds

Zhaoqiang Bai,[1,2] Lei Shen,[1,*] Guchang Han,[2] and Yuan Ping Feng[1,†]

[1] Department of Physics, National University of Singapore, 2 Science Drive 3, Singapore 117542, Singapore

[2] Data Storage Institute, Agency for Science, Technology and Research, 5 Engineering Drive 1, Singapore 117608, Singapore

**Abstract**

In the recent decade, the family of Heusler compounds has attracted tremendous scientific and technological interest in the field of spintronics. This is essentially due to their exceptional magnetic properties, which qualify them as promising functional materials in various data-storage devices, such as giant-magnetoresistance spin valves, magnetic tunnel junctions, and spin-transfer torque devices. In this article, we provide a comprehensive review on the applications of the Heusler family in magnetic data storage. In addition to their important roles in the performance improvement of these devices, we also try to point out the challenges as well as possible solutions, of the current Heusler-based devices. We hope that this review would spark further investigation efforts into efficient incorporation of this eminent family of materials into data storage applications by fully arousing their intrinsic potential.

1. Introduction

Ever since the discovery of giant magnetoresistance (GMR) by Fert and Grünberg (2007 Nobel Prize in Physics) in 1988,[1,2] intense research efforts, within the field of "spin electronics" or "spintronics",[3] have been boosted in both fundamental investigations and practical applications. Specifically, in the area of magnetic data storage, the GMR, by providing a sensitive and scalable read technique, has led to an increase of the areal recording density by more than two orders of magnitude (from ≈1 to ≈600 gigabits/ inch$^2$ in 2007).[3,4] In the early 1990s, the demonstration of GMR effect and oscillation behavior with respect to the thickness of non-magnetic (NM) layers was reported in the metallic superlattice systems Co/Cr, Co/Ru[5] and Co/Cu[6,7], using the economical sputtering technique. Subsequently, GMR was observed for the first time in a tri-layer spin-valve (SV) structure, composed of two ($Ni_{81}Fe_{19}$, $Ni_{80}Co_{20}$, Ni) ferromagnetic (FM) electrodes separated by (Cu, Ag, Au) NM spacers.[8] Later in 1997, SV-based current-in-plane (CIP) GMR sensor was commercialized by IBM as the read head of the hard disk drive (HDD), replacing the previous anisotropic magnetoresistance (AMR) technique. However, the planer geometry of the CIP SV sensors intrinsically limits further dimensional downscaling of the HDD read heads, and therefore, prevents higher storage density. In order to overcome this problem, the current-perpendicular-to-plane (CPP) configuration of GMR SVs was recently proposed as a promising architecture due to its geometrical compatibility with the shape of the read head track and intrinsic higher magnetoresistance values.[9,10]

Another big progress in spintronics related to data storage is the invention of the magnetic tunnel junction (MTJ), which is also a tri-layer stack composed of two FM layers sandwiching a NM insulator instead of metal. The story of the MTJ development began in the mid-1970s, when Jullière reported the observation of a small MR effect from a Fe/Ge/Co MTJ structure at low temperature.[11] However, the realization of remarkable and reproducible TMR effect had not been achieved until 1995, when the MTJ structure containing amorphous $Al_2O_3$ spacer was reported with a relatively large MR ratio ~70% at room temperature (RT).[12, 13] Afterwards, large TMR ratio was predicted in the MTJs with single-crystal MgO barrier by first-principles calculations.[14-16] Rather than the simple barrier model, the electron tunneling behavior of MgO-based MTJs is much more complicated, which is determined by the spin-dependent symmetry coupling of the transmission Bloch states between the FM electrodes and the NM MgO spacer. Subsequent experimental breakthrough was achieved in the MgO-based MTJ with much larger MR ratio ~200% at RT.[17, 18] In the recent years, the TMR effect was largely enhanced by the improved experimental techniques.[19] Although MTJs are believed as less favorable for HDD read heads due to the much larger intrinsic resistance-area product (RA), above $1\Omega cm^2$, compared with their all-metallic GMR counterparts,[10] they seems more eligible to be incorporated into the magnetic random access memory (MRAM).[4, 20] MRAM is claimed by its proponents to possess overwhelming advantages (nonvolatility, fast access and unlimited duration) over the current main-stream solid-state drive (SSD) memories, such as static random-access memory (SRAM), dynamic random-access memory (DRAM) and Flash, and would even become dominant over all types of data storage techniques as a "universal memory".[21] In reality, the first MRAM product, a 4-Mbit stand-along toggle memory, was commercialized in 2006 by Freescale.[22, 23] However, the current toggle MRAM may not be expected to scale well to small dimensions due to the intrinsic requirement of large write currents in the way of Oersted-field flipping. Such poor write-ability limits the number of elements that can be arrayed and degrades the layout efficiency of the memory. Moreover, large write currents also increase the power consumption well beyond that of SRAM or DRAM.[24]

The hope of breakthrough for the data writing was provided by the prediction in 1996 by Slonczewski and Berger that, instead of long-range effects mediated by the write current via its Oersted field, the magnetization orientation of a free magnetic layer could be controlled by a local means of manipulation via the transfer of spin angular momentum from a spin-polarized current, or in short, the spin-transfer torque (STT) effect.[25, 26] Specifically, a spin-polarized current of *s*-electrons are generated by transmission through or reflection from the thick reference layer and most of the electrons maintain this polarization as the current passes through the non-magnetic spacer. When the current approaches the thin free layer, however, an *s-d* exchange interaction occurs, which transfers the angular momentum from the polarized current to the free-layer magnetization, acting as an effective torque. This spin torque can oppose the intrinsic damping of the magnetic layer, reverse the direction of the magnetization and lead to a resistance change. The early experimental verification of the STT effect was made in Co/Cu multilayers[27] and Co/Cu/Co CPP-GMR SV nanopillars[28-30], whereas the interest was later shifted and focused on the association between STT and MRAM. In the spin-transfer torque random access memory (SPRAM), the switching threshold is determined by the injected spin-polarized current density, instead of the current, which makes it possible to ease many of the scaling limitations of the toggle MRAM.[31, 32] Besides, the elimination of the external write line would also lower the power consumption below that of both SRAM and DRAM.[24, 33, 34] Actually, several demonstrations of SPRAM have already been presented by industry,[35-37] which exhibit many aforementioned advantages, making it competitive as a future data-storage technique.[34]

For both HDD read heads and MRAM/SPRAM-based SSDs, large signal-to-noise ratio (SNR), which is related to the MR ratio of the GMR/TMR functional elements, is essential for the next-generation high-density data storage techniques.[3, 4, 38] A straightforward method of enhancing the MR is to use materials with high spin polarization (SP), such as FM or even half-metallic materials, for the electrodes of GMR/TMR devices.[3, 4, 38-40] Among all the possible materials, the family of Heusler alloys, first discovered by and named after Fritz Heusler in 1903,[41, 42] is widely believed as potential candidates eligible for constructing CPP-GMR and TMR

architectures.[43,44]

Heusler compounds are ternary inter-metallic face-centered cubic (fcc) crystals with the general formula XYZ (often called half-Heusler) or $X_2YZ$ (full-Heusler), in which X and Y are typically transition metals and Z is a main group element. The constituent elements of the Heusler compounds cover almost the whole periodic table, as shown in Fig. 1, which provides innumerable members in this family, and hence wide choices for the electronic structure tuning and material design of desirable properties, ranging from half-metallic ferromagnets (HMF),[39,45] completely-compensated ferrimagnets,[46] over nonmagnetic semiconductors,[47,48] to superconductors[49] and topological insulators.[50-54] A comprehensive description of the Heusler family can be found in Ref. 43. In this review article, however, we would focus on the applications of the Heusler compounds in the field of data storage.

Scientific interest in this field was sparked by the theoretical prediction[40,55] and the subsequent experimental verification[56] of the high Curie-temperature ($T_c$) half-metallicity in the bulk half-Heusler compound MnNiSb in the 1980s, which suggested the possibility of dramatic MR enhancement of the GMR/TMR devices by using the Heusler compounds as electrodes. However, the first MTJ with the MnNiSb epitaxial electrode yielded MR ratios as low as 9% at RT and 18% at low temperature,[57] respectively. This low MR ratio was attributed to the atomic-disorder which leads to the diminishing of the half-metallic gap around the Fermi level ($E_f$).[58] Similar low MR effect was reported for devices employing other half-Heusler compounds, for example, PtMnSb, in GMR SVs.[59, 60] Later on, research interest was shifted to Co-based full-Heusler compounds due to their expected larger MR effect, because they were shown to possess more stable half-metallicity, both in theory and experiment.[45, 61-67] The early successful demonstration of large MR values in the quaternary $Co_2Cr_{0.6}Fe_{0.4}Si$-based MTJs triggered enormous efforts focusing on the incorporation of the Co-based full Heusler compounds into both GMR and TMR devices,[66, 68] leading to a tremendous increase in the MR ratio during the recent decade.[69, 70]

In addition to the high spin polarization and high Curie temperature ($T_c$) mentioned above, Co-based Heusler compounds, such as $Co_2FeAl$ (CFA) and $Co_2Fe_xMn_{1-x}Si$ (CFMS), have a much lower saturation magnetization and damping constant compared with those of the conventional FM materials.[71, 72] This is of crucial importance to the reduction of the switching current and power consumption of the current STT devices. Remarkably, perpendicular magnetic anisotropy (PMA) has recently been reported in the CFA-based epitaxial stacking structures,[73-77] suggesting the feasibility of the application of CFA in the perpendicular FM electrodes of MTJs with high thermal stability at reduced dimension. Other than the well-known cubic full-Heusler compounds, a family of tetragonally-distorted Heusler compounds

$Mn_2YZ$ have recently emerged as another category of suitable materials due to their almost excellent fulfillment of the requirements for the electrodes of STT devices.[78-83]

This review article is organized as follows. In Section 2, a general description of the incorporation of Heusler compounds in MTJs is presented in Subsection 2.1, while in Subsection 2.2 we focus on the temperature dependence of the MR performance in these MTJs and summarize the possible ways in eliminating such an effect. The development of Heusler-based CPP-GMR devices is discussed in Section 3. Section 4 is dedicated to a novel all-Heusler design scheme for high-performance CPP-GMR and TMR junctions. An overview of the possible applications of Heusler compounds in STT devices is given in Sections 5, with Subsections 5.1 focusing on the cubic-phase compounds and 5.2 on the innovative tetragonally-distorted phase. Finally, a brief outlook of the future development of the Heusler-based data storage devices is addressed in Section 6 to conclude this review article.

## 2. Heusler Compound-based Magnetic Tunnel Junctions

### 2.1 Development

As mentioned above, a straightforward strategy to obtain a large TMR effect is to use electrode materials with an intrinsically high spin polarization. Following this consideration, a number of half-metallic (100% spin polarization) Heusler compounds have been widely studied and utilized, among which the first experimental observation of TMR in a B2-odered $Co_2Cr_{0.6}Fe_{0.4}Al(CCFA)$(10 nm)/$AlO_x$(1.8 nm)/CoFe(3 nm) MTJ was reported by Inomata *et al.* in 2003.[68] This Heusler-based MTJ was fabricated at RT on a thermally oxidized Si substrate without any buffer layers.[68] The MR ratio was 16% at RT and 26.5% at 5K, respectively. This important experiment pointed out the possibility of creating large TMR at RT in the Heusler-compound system. Further investigation showed that the $Co_2(Cr_{1-x}Fe_x)Al$ films possess the B2 or A2 structure, depending on the stoichiometric constitution, and the largest TMR value was obtained at *x*=0.4.[84, 85] Following this, intense experimental efforts have been made on the demonstration and enhancement of TMR in the MTJs with an amorphous alumina ($AlO_x$) barrier and various Co-based full-Heusler electrodes, e.g., CCFA,[84, 85] $Co_2MnAl$ (CMA),[86, 87] $Co_2MnSi$ (CMS),[88-90] $Co_2FeAl$ (CFA),[91] $Co_2FeSi$ (CFS),[92] $Co_2FeAl_{0.5}Si_{0.5}$ (CFAS)[93] and CMS/CFS multilayers.[94]

Recently, TMR ratio as high as 6000% was theoretically predicted in the Fe/MgO/Fe(001) MTJ from a ballistic conductance calculations.[14, 15] It was found that the decay rate of the wave function with the $\Delta_1$ symmetry is very slow as compared with that of the $\Delta_2$ and $\Delta_5$ channels in the single crystalline MgO barrier owing to the symmetry compatibility with the complex $\Delta_1$ band in the insulating gap of the MgO. Since bcc-Fe is half metallic on the $\Delta_1$ band, the Fe/MgO/Fe (001) MTJ can provide very large TMR. Three years later, high TMR ratios of about 180% in the Fe/MgO/Fe(001) MTJ[18] and about 220% in the CoFe/MgO/CoFe (001) MTJ[17] were successfully achieved at room temperature. Furthermore, the TMR ratio was raised to 604% at RT and 1144% at 5 K in the MTJ with CoFeB electrodes and a MgO barrier.[19] While the Fe/MgO/Fe(001) MTJ and related systems have succeeded in obtaining very large TMR ratio at RT, lower resistance is required in order for the MTJ to be compatible to other circuit elements to build cascade devices. To this end, one needs a tunneling junction with a thin MgO barrier. However, the TMR ratio of Fe/MgO/Fe (001) and related systems decreases rapidly as the thickness of the MgO barrier is reduced owing to the contribution of the minority-spin $\Delta_5$ and $\Delta_2$ channels to the tunneling conductance through a thin MgO barrier. The use of HMFs as electrodes in the MTJ with MgO barrier (HMF/MgO/HMF) can be expected to suppress the tunneling from the minority-spin $\Delta_5$ and $\Delta_2$ states in such a thin MgO barrier and has the potential to overcome this problem. The MTJs composed of the MgO barrier and half-metallic Co-based full Heusler compounds, e.g., $Co_2FeAl_{0.5}Si_{0.5}$,[95-97] $Co_2Cr_{0.6}Fe_{0.4}Al$,[98] and $Co_2MnSi$,[99-101] were fabricated in recent years, and the TMR ratio has been improved steadily, as summarized in Fig. 2. Furthermore, first-principles calculations suggested that the Co-based Heusler compounds, such as CMS, also have specific $\Delta_1$-band half-metallicity in addition to their total band half-metallicity, and might induce similar coherent tunneling and symmetry selective filtering effect[14, 16, 102] as that in (Co)Fe/MgO/(Co)Fe MTJs.[103] Remarkably, this proposal was later confirmed by a clear experimental observation of TMR oscillation behaviour with respect to the thickness of the MgO barrier, similar to that in Fe/MgO/Fe MTJs,[18, 104] even in the B2-disordered CFA/MgO/CFA MTJs.[105] However, it should be noted that high-quality MgO barriers are an indispensable factor for the achievement of such giant coherent tunneling effect; otherwise the effect would be destroyed even for the MTJs with L2$_1$ ordered CFA electrodes.[97]

### 2.2 The temperature effect

As can be seen in Fig. 2, although large TMR ratios have been achieved at low temperature in MTJs with Co-based full-Heusler compounds, these TMR values significantly decrease at elevated temperatures. To understand the origin of the TMR reduction at finite temperature, Lezaic *et al*. discussed the thermal collapse of the spin polarization by using an extended Heisenberg model together with *ab initio* calculation.[106] They found that the spin polarization of bulk CMS drops rapidly with increasing temperature for $T > 0.27T_c$, of which the decisive factor is the change in hybridization due to spin fluctuation. Furthermore, Chioncel *et al.* investigated the effects of electronic correlations in CMS by combining a theoretical dynamical mean field theory (DMFT) and an experimental tunneling-conductance spectroscopy measurement. They attributed the SP reduction at finite temperatures to the appearance of nonquasiparticle (NQS) states in the half-metallic gap of CMS, which extend across the Fermi level ($E_f$) and open an additional tunneling channel to the minority spin.[107] However, a further photoelectron spectroscopy measurement showed no distinct temperature dependence of the CMS valence-band electronic structure, which was in contrast with the NQS explanation.[108] Accordingly, the spin-mixing behavior, such as magnon excitations and inelastic scattering, was discussed as an alternative mechanism for the collapse of half-metallicity, and in turn, the rapid decrease of the TMR ratio at finite temperature.[53,109] These effects are due to the presence of interface states within the CMS minority gaps at the CMS/insulator junction. Moreover, Sakuma et al. showed, based on their first-principles calculations, that the exchange constant of interfacial Co spin moments at the CMS/MgO(001) and CMA/MgO interfaces is relatively small compared to that of bulk CMS and CMA, leading to instability of the interfacial Co spin moments at finite temperature and the strong temperature dependence of the TMR ratio.[110] Miura et al. further investigated the effects of spin-flip scattering on the TMR value in a CMS/MgO/CMS MTJ on the basis of the first-principles calculations.[111] They confirm that the non-collinear magnetic structures of interfacial Co spin moments, resulting from the thermal fluctuations, cause the spin-flip scattering and lead to a significant reduction of the TMR.

The stabilization of half-metallicity of the Heusler electrodes is one of the possible solutions for minimizing such temperature effect and achieving high TMR ratio at RT. The substitutional series of quaternary Heusler compounds, $Co_2FeAl_{1-x}Si_x$, were proposed as promising candidates because of their localized moments at the Fe sites, simplicity of majority bands in the vicinity of the Fermi level, and most importantly, their tunable Fermi level within the minority gap.[96, 112, 113] It was shown by first principles calculations that the whole series of quaternary $Co_2FeAl_{1-x}Si_x$ compounds exhibits HMF character and an almost linear concentration factor (x) dependence of the Fermi level position relative to the half-metallic gap. Among the series, $Co_2FeAl_{0.5}Si_{0.5}$ (x=0.5) has its Fermi level located right in the middle of the minority gap.[112] Such a median position of $E_f$ would efficiently enhance the thermal stability of the half-metallicity because it is far from the edges of both valence and conduction bands, and therefore, reduces the sensitivity to various thermal fluctuation effects.[112] This Fermi level tuning behavior was experimentally demonstrated subsequently by the differential conductance measurements in a CFAS/$(MgAl_2)O_x$/CoFe MTJ, showing the highest effective spin polarization ($P_{eff}$) of up to 0.91, and the weakest temperature dependence of the $P_{eff}$ among all known half metals.[114] Further study in the same work showed that the decay of $P_{eff}$ follows $T^{3/2}$ law, $P_{eff}(T)=P_0(1-\alpha T^{3/2})$, perfectly, which indicates that the depolarization of CFAS is determined by spin wave excitation only. However, MTJs with CFAS electrodes and the MgO spacer, fabricated either by magnetron sputtering[96] or molecular beam epitaxy[115], still exhibited large temperature dependence of TMR values.

Moreover, the temperature dependence of TMR is also largely affected by the morphology and quality of the interfaces between the half-metallic electrodes and the spacer. It was estimated that the decay factor α is larger at the interface than that in the bulk, indicating comparatively larger interfacial thermal fluctuation effect.[114] A further investigation into the fully epitaxial B2-ordered CFA/MgO/CFA MTJs shows that the structural quality of the MgO/top-CFA interface is heavily reduced compared with that of the bottom-CFA/MgO junction.[116] If assuming the same morphology of the upper MgO/top-CFA interface as that of the MgO-buffer/CFA junction, the predicted MR ratio could be as high as 1800% at RT. Therefore, it should be worthwhile to make further experimental efforts to realize high-quality MgO/top-Heusler electrode interface, although it is a challenging issue based on current technology.

### 3. Heusler-based CPP-GMR Spin Valves

Continuous evolution of HDD read heads with higher sensor output, lower resistance, and a higher bit resolution is required for the further ultrahigh density magnetic recording. Low resistance MR devices are of urgent requirement for impedance matching between read sensors and the preamplifiers, for lower electric noises, and for high frequency data transfer.[117-119] A read-sensor RA less than 0.1 $\Omega\mu m^2$, at least, is required for the recording densities exceeding 2 terabits/inch$^2$.[119] This is a big challenge for the currently used MTJs with a high impedance insulator spacer, but can be easily achieved for the CPP-GMR SVs composed of all-metallic layers, whose RA values are typically below $0.05\Omega\mu m^2$. However, the SNR of the CPP-GMR SVs with conventional ferromagnetic (FM) materials such as the CoFe alloy, i.e., resistance change-area product (ΔRA) of ~1 mΩ μm$^2$,[120, 121] must be improved substantially. Similar to the MTJ case, the utilization of highly spin-polarized FM materials such as Co-based full-Heusler compounds is expected to provide large spin-dependent scatterings both in the FM layers and at the interfaces between the FM and the spacer layers, thereby improving the ΔRA of the CPP-GMR SVs.[70]

The first Co-based Heusler CPP-GMR device was achieved in 2006 consisting of two Co$_2$MnSi (CMS) electrodes separated by a 3 nm Cr spacer.[122] A ΔRA of 19 mΩ μm$^2$ was reported, which is approximately 10 times larger than that of the conventional tri-layer system, such as CoFe/Cu (< 2 mΩ μm$^2$),[120, 121] suggesting that CPP-GMR with a high-quality Co-based half-metallic full-Heusler electrode has a great potential for the HDD read heads. However, the maximum MR ratio was only 2.4% at room temperature, and increased slightly to 5.2% by improving the L2$_1$ ordering of the CMS electrodes via proper annealing.[123]

It should be noted that, besides the electrode materials, the choice of the spacer layer is also an important issue, since an epitaxial growth of the Heusler thin film on the spacer material is required to form fully epitaxial Heusler/NM spacer/Heusler tri-layers. Moreover, according to the Valet and Fert Model for CPP-GMR architecture,[9] ΔRA is determined by the intrinsic spin-asymmetry coefficients not only in the bulk FM electrode (β) but also at the FM/NM interface (γ). Good band matching (Fermi surface matching) between the electrode and spacer materials for the majority spin is a predominating factor in achieving large SNR.[82, 124-128] Besides, a large spin-diffusion length (SDL) and low resistivity are also necessary for the promising spacer materials.

These considerations, combined with a small lattice mismatch, led to the selection of silver as an ideal spacer material coupled with the Heusler electrodes. Consequently, a CPP-GMR ratio of 6.9% at room temperature was realized for a $Co_2FeAl_{0.5}Si_{0.5}/Ag/Co_2FeAl_{0.5}Si_{0.5}$ tri-layer stack.[129] Further enhanced CPP-GMR ratios of 34% for the same system[130] and of 24% in an antiferromagnetic interlayer-exchange-coupling (AFM-IEC) architecture[131] were reported. For CMS-based CPP-GMR stacks, much larger MR ratios of 28.8%[132] and 36.4%[133] were also observed by the substitution of Cr spacers by the Ag ones. In the last few years, the employment of various quaternary Co-based Heusler compounds as electrodes were also witnessed with relatively large MR values: 8.8%[134] and 10.2%[135] of $Co_2MnGa_{0.5}Sn_{0.5}$, 41.7% of $Co_2FeGa_{0.5}Ge_{0.5}$,[136] and 74.8%, the largest value so far to our best knowledge, of $Co_2Fe_{0.4}Mn_{0.6}Si$.[137] The development of MR values in CPP-GMR SVs with various electrode and spacer materials is illustrated in Fig. 3. It is worth to mention that, generally, there are two main advantages for using the well-designed quaternary Heusler compounds, of which the first one is, as mentioned in Section 3, the enhancement of thermal stability by tuning $E_f$ to the middle of the minority gap through varying the composition ratio. The second advantage is that the generated quaternary compounds can inherit the merits while avoid the shortcomings of its two parent ternary compounds by balancing their magnetic and spin states. For example, $Co_2(Cr, Fe)Al$ shares the modestly high SP of $Co_2CrAl$ and high $T_c$ of $Co_2FeAl$ via mixing them together. [66, 67]

Attempts were also made to find other potential materials, such as Cu, as the spacer. Generally, however, the reported MR values were not as large as those stacks with Ag spacers.[138-142] The reason was possibly the larger lattice mismatch between Cu and Heusler, which resulted in the lower degree of structural $L2_1$ ordering of the top Heusler layers on the Cu spacer and the large amount of twins in the Cu layers.[141]

Narrow CPP-GMR read heads, incorporating Heusler materials as reference layers, were successfully tested using a conventional spin-stand system. Thus, the capability of the CPP-GMR technology for ultra-high density magnetic recording was demonstrated. Further investigation and exploration of Heusler-compound systems based on thermodynamics and the electronic structure, however, is necessary to the development of CPP-GMR tag materials and makes the heads superior to the TMR ones.[139] In addition, searching for new materials for the spacer layer is another

important issue in views of the lattice- and band-matching with the Heusler alloy FM layers.[70]

## 4. All-Heusler TMR/GMR junctions

In order to push the CPP-GMR values into the applicable range, an alternative approach, namely "all-Heusler" architecture, to improving the lattice- and band-matching for the optimization of the interface scattering properties, was proposed in $Co_2MnGe(Si)/Ru_2CuSn/Co_2MnGe(Si)$ trilayer stacks.[125, 143] However, the measured MR ratio was only ~7%, which could possibly be attributed to the reduced interfacial spin polarization caused by the large lattice mismatch, atomic disorder and surface states in the MnGe terminated $Co_2MnGe/Ru_2CuSn$ interface. Recently, an interesting all-Heusler compound interface $Co_2CrSi/Cu_2CrAl$ was proposed.[124, 128] $Co_2CrSi$ and $Cu_2CrAl$ have the same lattice structure and their lattice constants match very well. First-principles electronic structure and transport calculations predicted an enhanced spin scattering asymmetry at the $Co_2CrSi/Cu_2CrAl$ interface, which makes the $Co_2CrSi/Cu_2CrAl/Co_2CrSi$ stacking structure as an appealing candidate for the all-Heusler CPP-GMR architecture. Other all-Heusler interfaces like $L2_1$ $Co_2MnSi/Ni_2NiSi$,[126] $Fe_2CrSi/Cu_2CrAl$,[144] as well as $Cl_b$ NiMnSb/NiCuSb,[126] has also been proposed by first-principles calculation. By taking advantage of the intrinsically matched energy bands and Fermi surfaces among a number of full-Heusler compounds, Bai et al. suggested a design scheme of high-performance all-Heusler CPP-GMR junctions with large spin-asymmetry, high spin-filter efficiency, and consequently high MR ratio.[145] Based on consideration of fulfilling the requirement of the interfacial structural and chemical compatibilities, Chadov *et al.* proposed a general rule of material selection for the all-Heusler scheme,[146] which intends to provide stable high spin polarization at the interfaces of the magnetoresistance junctions. This can be realized by joining the semiconducting and half-metallic Heusler materials with similar structures. It was verified by first-principles calculations that the interface remains half-metallic if the nearest interface layers effectively form a stable Heusler material with the properties intermediately between the surrounding bulk parts.

## 5. Heusler compounds with perpendicular magnetic anisotropy

### 5.1 Cubic phase PMA Heusler

The magnetoresistance phenomena discussed in the previous section (GMR or TMR) allows controlling an electron flow through a magnetic nanostructure by its magnetic state. The reciprocal phenomenon also exists. Spin-transfer torque (STT), which was predicted independently by Slonczewski and Berger in 1996,[25, 26] is an attractive spintronics phenomenon in which spin-polarized current flowing between FM layers can change the relative alignment of their magnetization orientations. During the recent years, considerable interest has been witnessed for the development of STT-based magnetic random access memory due to the more precise addressing and lower energy consumption compared with its traditional counterparts, which is necessary when the size of the transistor shrinks to sub-100 nm regime.[33, 147-153] Currently, the investigations of spin-transfer switching in MTJs, which are the key element of the SPRAMs, have mainly been focused on the stacking structure containing a CoFeB/MgO/CoFeB tri-layer due to its large MR ratio.[149-152] However, the switching current density of these MTJs is of the order of $MA/cm^2$, which is too high for practical application. In fact, the dilemma of balancing the writability and the thermal stability becomes a crucial issue as the size of SPRAMs is shrinking[154]: for a smaller MTJ size, a higher writing current density is required to overcome the increased energy barrier for maintaining the thermal stability; however, only a large transistor can supply such a high current due to the current density limitation, thus the high storage density cannot be achieved for such one transistor and one memory design scheme. In other words, the main challenge for implementing SPRAM in high-density and high-speed memory is to reduce substantially the intrinsic current density required to switch the FM magnetization direction while maintaining high thermal stability required for long-term data retention.

A few years ago, an elegant scheme of perpendicular-to-plane polarizer (PERP), in which the FM electrodes of the STT device are composed of perpendicular magnetic anisotropy materials, was proposed as a potential alternative to the conventional in-plane longitudinal polarizer (LONG) to overcome the challenge mentioned above.[155-158] Two advantages can be achieved by using such PMA electrodes with magnetic anisotropy normal to the film surface. On one hand, for the patterned device, the magnetization of PMA materials is more uniform and does not suffer from the thermal instability due to magnetization curling observed at the edge of in-plane case, which would be beneficial to the reduction of the aspect ratio (length/width) of the film and hence the bit size.[155] On the other hand, the switching current of the MTJ can be decreased by introducing PMA free layer, to counteract the effect of the large out-of-plane demagnetizing field which inhabits current-induced switching.[24, 31, 159-162]

Various PMA materials, including Co/Ni[163] and (Co,Fe)/(Pt,Pd) multilayers,[164, 165] $L1_0$-ordered FePt[166] and CoPt[167] alloys, and rare-earth/transition-metal alloys[155, 168] have been investigated for use as the PMA electrodes for perpendicular MTJs (p-MTJs) because of their high perpendicular anisotropy energy, resulting in high thermal stability. However, the low spin polarization of these conventional PMA materials and the large lattice mismatch with an MgO barrier reduce the TMR ratio of MTJs. They are also inclined to exhibit a large damping constant($\alpha$), e.g., $\alpha_{L1_0\text{-FePt}} \geq$ 0.055 due to the strong spin-orbit coupling,[169] and hence, high current density ($J_{c0}$) for STT switching. Recently, perpendicular magnetization of ultrathin films of soft magnetic materials such as Fe and fcc-Co have been realized when sandwiched by Pt/Pd and oxides (amorphous $AlO_x$ and crystalline MgO layers), which are attributed to the hybridization of orbitals between ferromagnetic and oxygen atoms and the spin-orbit coupling.[170, 171] Remarkably, the perpendicularly magnetized CoFeB/MgO/CoFeB MTJs were demonstrated to be effective for achieving high thermal stability, high TMR, and low critical current density simultaneously in an MTJ nanopillar, in which a large TMR of 124% and $J_{c0} = 3.9 \times 10^6$ A/cm$^2$ have been obtained.[153]

In order to further improve the SNR and reduce the switching current density, PMA thin films with a large tunneling spin polarization $P$ such as half-metallic Heusler compounds are desired for the electrodes in MTJs. Among all the possible Heusler candidates, highly spin-polarized B2-ordered Co$_2$FeAl possesses the smallest Gilbert damping constant (≈0.001 after post-annealing at 600 ℃),[71] which is a substantial factor in reducing the STT switching current. Combined with its coherent tunneling properties as demonstrated in the CFA/MgO/CoFe MTJs,[105] CFA is regarded as a promising material for the STT-devices, and therefore, has been intensely investigated. Wang *et al*. first demonstrated the PMA in CFA/Pt multilayers prepared on Pt buffered MgO(001) substrates by magnetron sputtering, of which the PMA energy density $K_u$ is estimated to be $1.45 \times 10^6$ erg/cm$^3$ at room temperature, which is comparable to that of the Co/(Pd,Pt) multilayers, and therefore, suggests the possibility of the incorporation of the highly spin-polarized CFA film into p-MTJ structures. However, the PMA as demonstrated is not only highly sensitive to the number of periods *n* of the multilayers but also confined to very thin CFA film (~0.6 nm).[74] In order to achieve thicker perpendicularly magnetized Heusler films, a sandwiched structure Pt/CFA/MgO was proposed by Li *et al.*, claiming that the PMA is thermally stable up to 2 nm of the CFA film with $K_u$ $1.3\times10^6$ erg/cm$^3$.[76] Subsequently, Wen *et al.* reported demonstration of PMA in the structures of $Co_2FeAl$/MgO, and inverse MgO/$Co_2FeAl$, with $K_u$ of 2-3×10$^6$ erg/cm$^3$.[77] Other ways of the fabricating perpendicular magnetized Heusler films have also been witnessed, e.g. the co-sputtering of CFA with terbium.[73, 75]

At the device level, Sukegawa *et al.* experimentally demonstrated the spin-transfer switching in the epitaxial CFA-based MTJs CFA/MgO/CoFe prepared on a Cr buffer layer. However, the switching current density is still quite large (~29 MA/cm$^2$) due to enhancement of the Gilbert damping factor of the CFA by the Cr buffer.[172] Wen *et al.* fabricated ultrathin-$Co_2FeAl$/MgO/$Co_{20}Fe_{60}B_{20}$ p-MTJs, in which an out-of-plane TMR of 53% was achieved at RT and could be further enhanced to 91% by the improvement of the interface quality via inserting a 0.1-nm-thick Fe layer between the $Co_{20}Fe_{60}B_{20}$ electrode and MgO spacer.[173] Additional enhancement of the TMR ratio could be obtained by further improvement of the B2 ordering of the $Co_2FeAl$ electrode, and the interface structure.

### 5.2 Tetragonally distorted phase PMA Heusler

In addition to the cubic Heusler compounds, a group of tetragonally distorted Heusler compounds with the inverse structure, *DO*$_{22}$-phase $Mn_2YZ$, have recently attracted great scientific interest in the field of spin-transfer torque applications. In this structure, the Mn atoms occupy two different sites, forming one tetragonal and one octahedral sublattice. Due to the crystal field theory, Jahn-Teller distortion occurs in the tetragonal sites to minimize the total energy of $Mn^{3+}$ $d^4$ electronic configuration, which elongates the lattice along c-axis and reduces the crystal symmetry from the cubic $F\bar{4}3M$ to the tetragonal I4/mmm group.

As a prototype, the compensated ferrimagnet $Mn_3Ga$ has been thoroughly studied. It turns out that the binary compound possesses a magnetization easy axis pointing perpendicular to the thin film surfaces ($K_u^{eff} = 1.2\times10^7$ erg/cm$^3$), which is of crucial importance to maintain high thermal stability at moderate coercive fields for long-term data retention in high-density magnetic data-storage devices.[78-80, 83, 153] In addition, the combination of high spin polarization, high Curie temperature, low magnetic damping constant and low saturation magnetization ($M_s$ = 250 emu/cm$^3$) almost fulfills all the prerequisites for low energy-consumption spin-transfer torque devices.[78-80, 83, 174] Recent experimental efforts have been made on the demonstration of tunneling magnetoresistance effect within the epitaxially grown magnetic tunnel junctions using $Mn_{3-x}Ga$ compounds as electrodes.[82, 175, 176] However, the MR ratio is very small and far below the application range. Based on first-principles calculations, Bai *et al.* concluded that the low MR is due to the poor wavefunction symmetry matching between the $Mn_{3-x}Ga$ (x ~ 0) electrodes and the MgO spacer.[177] Further experimental effort is highly demanded in order to push the MR ratio to the application range.

Searching for new tetragonal distorted $Mn_2YZ$ (Y is typically transition metals more electronegative that Mn) Heusler materials, for instance $Mn_{3-x}Co_xGa$[178, 179] and $Mn_2PtIn$[180], which have suitably designed properties as new magnetic layers in spin-torque devices, is an active field of ongoing research. A valence rule proposed by Wurmehl *et al*,[46] which combines the well-known Slater–Pauling rule[181, 182] with the Kübler rule[45], provides a guide for finding new compounds with half-metallic-type behavior and completely compensated moments. On one hand, Slater–Pauling rule determines the total magnetic moment from the mean number of valence electrons. According to this rule, the half-metallic system of interest must have 24 valence electrons. The Kübler rule, on the other hand, states that Mn in Heusler compounds always carries a high local magnetic moment at the octahedral site (Y site of the formula $X_2YZ$). Furthermore, it was found that even partial occupation of the Y position by Mn is sufficient to enforce a local magnetic moment on this site. Noticeably, very recently, a design scheme of novel tetragonal Heusler compounds based on the ab initio electronic structure calculation has been proposed.[183] The band Jahn-Teller instability, which leads to the tetragonal distortion to the unstable cubic $Mn_2XY$, can be characterized by the van Hove singularities (high peaks of density of states) in proximity of the Fermi energy.[184, 185]

## 6. Summary and Outlook

In this article, we give a comprehensive review of the Heusler family with special focus on its broad applications in the field of magnetic data storage, ranging from CPP-GMR read heads, to MRAM arrays, and to the very recently emerging SPRAMs. The overwhelming advantages over the conventional FM materials lead to the superiority of the Heusler compounds as outstanding functional building blocks for these spintronics devices.

Further investigation and exploration of the Heusler compound systems based on the understanding of their thermodynamic and the electronic properties is highly demanded for the performance improvement of the current Heusler-based data storage devices. The search for new promising Heusler compounds will also be a fruitful research area in the coming years.

Remarkably, a new class of half-Heusler compounds has recently been introduced as topological insulators (TIs),[43, 44, 50-54, 186-197] which are insulating bulk covered by symmetry-protected metallic surfaces.[198, 199] The most important characteristic of TIs is the band inversion behavior due to the strong spin-orbital coupling (SOC). As summarized in Fig. 5, tens of Heusler compounds have been predicted to show such character, analogue to the binary compound HgTe, a prototypical TI.[51, 53] It has been demonstrated that both strain and electric/magnetic fields can be used to modify the electronic structure (band gap) and transport (carry mobility) properties of TIs.[200-203] Because of such unique electronic and magnetic features, TIs show potential developments in the field of spintronics and data storage application. For example, very recently, a topological insulating GeTe/Sb$_2$Te$_3$ superlattice was proposed for the phase-change random access memory (PCRAM).[204] In consideration of the wide choices provided the Heusler family, it is natural to assume that this functionality could be transferrable to the Heusler-based TI superlattice architecture with even larger space for property tuning due to their ternary composition. Such novel discoveries, as well as the established promising magnetic properties as reviewed, would make the Heusler family continue to play a key role in the future data storage area.

**Acknowledgement**:
This work is supported by the Singapore Agency for Science, Technology, and Research (A*STAR) grant (SERC Grant Nos. 0921560121 and 0721330044).

**Corresponding authors**:
shenlei@nus.edu.sg; phyfyp@nus.edu.sg

# Reference

1. M. N. Baibich, J. M. Broto, A. Fert, F. N. Van Dau, F. Petroff, P. Etienne, G. Creuzet, A. Friederich and J. Chazelas, Physical Review Letters **61** (21), 2472 (1988).
2. G. Binasch, P. Grünberg, F. Saurenbach and W. Zinn, Physical Review B **39** (7), 4828 (1989).
3. A. Fert, Reviews of Modern Physics **80** (4), 1517 (2008).
4. C. Chappert, A. Fert and F. N. Van Dau, Nature Materials **6** (11), 813 (2007).
5. S. S. P. Parkin, N. More and K. P. Roche, Physical Review Letters **64** (19), 2304 (1990).
6. D. H. Mosca, F. Petroff, A. Fert, P. A. Schroeder, W. P. Pratt Jr and R. Laloee, Journal of Magnetism and Magnetic Materials **94** (1–2), L1-L5 (1991).
7. S. S. P. Parkin, R. Bhadra and K. P. Roche, Physical Review Letters **66** (16), 2152 (1991).
8. B. Dieny, V. S. Speriosu, S. S. P. Parkin, B. A. Gurney, D. R. Wilhoit and D. Mauri, Physical Review B **43** (1), 1297 (1991).
9. T. Valet and A. Fert, Physical Review B **48** (10), 7099 (1993).
10. M. Takagishi, K. Koi, M. Yoshikawa, T. Funayama, H. Iwasaki and M. Sahashi, IEEE Transactions on Magnetics **38** (5), 2277 (2002).
11. M. Julliere, Physics Letters A **54** (3), 225 (1975).
12. T. Miyazaki and N. Tezuka, Journal of Magnetism and Magnetic Materials **139** (3), L231-L234 (1995).
13. J. S. Moodera, L. R. Kinder, T. M. Wong and R. Meservey, Physical Review Letters **74** (16), 3273 (1995).
14. W. H. Butler, X. G. Zhang, T. C. Schulthess and J. M. MacLaren, Phys Rev B **63** (5), 054416 (2001).
15. J. Mathon and A. Umerski, Phys Rev B **63** (22), 220403 (2001).
16. X. G. Zhang and W. H. Butler, Physical Review B **70** (17), 172407 (2004).
17. S. S. P. Parkin, C. Kaiser, A. Panchula, P. M. Rice, B. Hughes, M. Samant and S.-H. Yang, Nat Mater **3** (12), 862 (2004).
18. S. Yuasa, T. Nagahama, A. Fukushima, Y. Suzuki and K. Ando, Nat Mater **3** (12), 868 (2004).
19. S. Ikeda, J. Hayakawa, Y. Ashizawa, Y. M. Lee, K. Miura, H. Hasegawa, M. Tsunoda, F. Matsukura and H. Ohno, Applied Physics Letters **93** (8), 082508 (2008).
20. J. M. Daughton, Journal of Applied Physics **81** (8), 3758 (1997).
21. J. Åkerman, Science **308** (5721), 508 (2005).
22. M. Durlam, D. Addie, J. Akerman, B. Butcher, P. Brown, J. Chan, M. DeHerrera, B. N. Engel, B. Feil, G. Grynkewich, J. Janesky, M. Johnson, K. Kyler, J. Molla, J. Martin, K. Nagel, J. Ren, N. D. Rizzo, T. Rodriguez, L. Savtchenko, J. Salter, J. M. Slaughter, K. Smith, J. J. Sun, M. Lien, K. Papworth, P. Shah, W. Qin, R. Williams, L. Wise and S. Tehrani, presented at the Electron Devices Meeting, 2003. IEDM '03 Technical Digest. IEEE International, 2003 (unpublished).
23. B. N. Engel, J. Akerman, B. Butcher, R. W. Dave, M. DeHerrera, M. Durlam, G. Grynkewich, J. Janesky, S. V. Pietambaram, N. D. Rizzo, J. M. Slaughter, K. Smith, J. J. Sun and S. Tehrani, IEEE Transactions on Magnetics **41** (1), 132 (2005).
24. J. A. Katine and E. E. Fullerton, Journal of Magnetism and Magnetic Materials **320** (7), 1217 (2008).
25. L. Berger, Physical Review B **54** (13), 9353 (1996).
26. J. C. Slonczewski, Journal of Magnetism and Magnetic Materials **159** (1–2), L1-L7 (1996).

27. M. Tsoi, A. G. M. Jansen, J. Bass, W. C. Chiang, M. Seck, V. Tsoi and P. Wyder, Physical Review Letters **80** (19), 4281 (1998).
28. F. J. Albert, J. A. Katine, R. A. Buhrman and D. C. Ralph, Applied Physics Letters **77** (23), 3809 (2000).
29. J. A. Katine, F. J. Albert, R. A. Buhrman, E. B. Myers and D. C. Ralph, Physical Review Letters **84** (14), 3149 (2000).
30. E. B. Myers, D. C. Ralph, J. A. Katine, R. N. Louie and R. A. Buhrman, Science **285** (5429), 867 (1999).
31. J. Z. Sun, Physical Review B **62** (1), 570 (2000).
32. J. C. Slonczewski, Physical Review B **71** (2), 024411 (2005).
33. E. Chen, D. Apalkov, Z. Diao, A. Driskill-Smith, D. Druist, D. Lottis, V. Nikitin, X. Tang, S. Watts, S. Wang, S. A. Wolf, A. W. Ghosh, J. W. Lu, S. J. Poon, M. Stan, W. H. Butler, S. Gupta, C. K. A. Mewes, T. Mewes and P. B. Visscher, IEEE Transactions on Magnetics **46** (6), 1873 (2010).
34. T. Kawahara, K. Ito, R. Takemura and H. Ohno, Microelectronics Reliability **52** (4), 613 (2012).
35. M. Hosomi, H. Yamagishi, T. Yamamoto, K. Bessho, Y. Higo, K. Yamane, H. Yamada, M. Shoji, H. Hachino, C. Fukumoto, H. Nagao and H. Kano, presented at the Electron Devices Meeting, 2005. IEDM Technical Digest. IEEE International, 2005 (unpublished).
36. T. Kawahara, R. Takemura, K. Miura, J. Hayakawa, S. Ikeda, Y. M. Lee, R. Sasaki, Y. Goto, K. Ito, T. Meguro, F. Matsukura, H. Takahashi, H. Matsuoka and H. Ohno, IEEE Journal of Solid-State Circuit **43** (1), 109 (2008).
37. R. Takemura, T. Kawahara, K. Miura, H. Yamamoto, J. Hayakawa, N. Matsuzaki, K. Ono, M. Yamanouchi, K. Ito, H. Takahashi, S. Ikeda, H. Hasegawa, H. Matsuoka and H. Ohno, IEEE Journal of Solid-State Circuit **45** (4), 869 (2010).
38. P. A. Grünberg, Reviews of Modern Physics **80** (4), 1531 (2008).
39. C. Felser, G. H. Fecher and B. Balke, Angewandte Chemie International Edition **46** (5), 668 (2007).
40. R. A. de Groot, F. M. Mueller, P. G. v. Engen and K. H. J. Buschow, Physical Review Letters **50** (25), 2024 (1983).
41. W. S. F. Heusler, and E. Haupt, Verh. Deutsche Physikalische Gesellschaft **5** (1903).
42. F. Heusler, Verh. Deutsche Physikalische Gesellschaft **5**, 219 (1903).
43. T. Graf, C. Felser and S. S. P. Parkin, Progress in Solid State Chemistry **39** (1), 1 (2011).
44. T. Graf, S. S. P. Parkin and C. Felser, IEEE Transactions on Magnetics **47** (2), 367 (2011).
45. J. Kübler, A. R. William and C. B. Sommers, Physical Review B **28** (4), 1745 (1983).
46. S. Wurmehl, H. C. Kandpal, G. H. Fecher and C. Felser, Journal of Physics: Condensed Matter **18** (27), 6171 (2006).
47. J. Pierre, R. Skolozdra, J. Tobola, S. Kaprzyk, C. Hordequin, M. A. Kouacou, I. Karla, R. Currat and E. Lelievre-Berna, Journal of Alloys and Compounds. **262**, 101 (1997).
48. D. Jung, H. J. Koo and M. H. Whangbo, Journal of Molecular Structure: THEOCHEM **527**, 113 (2000).
49. J. Winterlik, G. H. Fecher, A. Thomas and C. Felser, Physical Review B **79** (6), 064508 (2009).
50. W. Al-Sawai, H. Lin, R. S. Markiewicz, L. A. Wray, Y. Xia, S. Y. Xu, M. Z. Hasan and A. Bansil, Physical Review B **82** (12), 125208 (2010).
51. S. Chadov, X. L. Qi, J. Kubler, G. H. Fecher, C. Felser and S. C. Zhang, Nature Materials **9** (7), 541 (2010).

52. W. Feng, D. Xiao, Y. Zhang and Y. Yao, Physical Review B **82** (23), 235121 (2010).
53. H. Lin, L. A. Wray, Y. Q. Xia, S. Y. Xu, S. A. Jia, R. J. Cava, A. Bansil and M. Z. Hasan, Nature Materials **9** (7), 546 (2010).
54. D. Xiao, Y. Yao, W. Feng, J. Wen, W. Zhu, X.-Q. Chen, G. M. Stocks and Z. Zhang, Physical Review Letters **105** (9), 096404 (2010).
55. K. E. H. M. Hanssen and P. E. Mijnarends, Physical Review B **34** (8), 5009 (1986).
56. K. E. H. M. Hanssen, P. E. Mijnarends, L. P. L. M. Rabou and K. H. J. Buschow, Physical Review B **42** (3), 1533 (1990).
57. C. T. Tanaka, J. Nowak and J. S. Moodera, Journal of Applied Physics **86** (11), 6239 (1999).
58. S. J. Jenkins, Physical Review B **70** (24), 245401 (2004).
59. M. C. Kautzky, F. B. Mancoff, J. F. Bobo, P. R. Johnson, R. L. White and B. M. Clemens, Journal of Applied Physics **81** (8), 4026 (1997).
60. P. R. Johnson, M. C. Kautzky, F. B. Mancoff, R. Kondo, B. M. Clemens and R. L. White, IEEE Transactions on Magnetics **32** (5), 4615 (1996).
61. S. Ishida, S. Fujii, S. Kashiwagi and S. Asano, Journal of Physical Society of Japan **64** (6), 2152 (1995).
62. W. E. Pickett and J. S. Moodera, Physics Today **54** (5), 39 (2001).
63. I. Galanakis, P. H. Dederichs and N. Papanikolaou, Physical Review B **66** (17), 174429 (2002).
64. S. Picozzi, A. Continenza and A. J. Freeman, Physical Review B **66** (9), 094421 (2002).
65. N. Auth, G. Jakob, T. Block and C. Felser, Physical Review B **68** (2), 024403 (2003).
66. T. Block, C. Felser, G. Jakob, J. Ensling, B. Muhling, P. Gutlich and R. J. Cava, Journal of Solid State Chemistry **176** (2), 646 (2003).
67. C. Felser, B. Heitkamp, F. Kronast, D. Schmitz, S. Cramm, H. A. Durr, H. J. Elmers, G. H. Fecher, S. Wurmehl, T. Block, D. Valdaitsev, S. A. Nepijko, A. Gloskovskii, G. Jakob, G. Schonhense and W. Eberhardt, Journal of Physics: Condensed Matter **15** (41), 7019 (2003).
68. K. Inomata, S. Okamura, R. Goto and N. Tezuka, Japanese Journal of Applied Physics **42**, L419 (2003).
69. K. Inomata, N. Ikeda, N. Tezuka, R. Goto, S. Sugimoto, M. Wojcik and E. Jedryka, Science and Technology of Advanced Materials **9** (1), 014101 (2008).
70. T. M. Nakatani, N. Hase, H. S. Goripati, Y. K. Takahashi, T. Furubayashi and K. Hono, IEEE Transactions on Magnetics. **48** (5), 1751 (2012).
71. S. Mizukami, D. Watanabe, M. Oogane, Y. Ando, Y. Miura, M. Shirai and T. Miyazaki, Journal of Applied Physics **105** (7), 07D306 (2009).
72. M. Oogane, T. Kubota, Y. Kota, S. Mizukami, H. Naganuma, A. Sakuma and Y. Ando, Applied Physics Letters **96** (25), 252501 (2010).
73. X. Q. Li, X. G. Xu, D. L. Zhang, J. Miao, Q. Zhan, M. B. A. Jalil, G. H. Yu and Y. Jiang, Applied Physics Letters **96** (14), 142505 (2010).
74. W. H. Wang, H. Sukegawa and K. Inomata, Applied Physical Express **3** (9), 093002 (2010).
75. X. Q. Li, X. G. Xu, S. Q. Yin, D. L. Zhang, J. Miao and Y. Jiang, Journal of Magnetism and Magnetic Materials **323** (14), 1914 (2011).
76. X. Q. Li, S. Q. Yin, Y. P. Liu, D. L. Zhang, X. G. Xu, J. Miao and Y. Jiang, Applied Physical Express **4** (4), 043006 (2011).
77. Z. C. Wen, H. Sukegawa, S. Mitani and K. Inomata, Applied Physical Letters **98** (24), 192505 (2011).

78. B. Balke, G. H. Fecher, J. Winterlik and C. Felser, Applied Physics Letters **90** (15), 152504 (2007).
79. J. Winterlik, B. Balke, G. H. Fecher, C. Felser, M. C. M. Alves, F. Bernardi and J. Morais, Physical Review B **77** (5), 054406 (2008).
80. F. Wu, S. Mizukami, D. Watanabe, H. Naganuma, M. Oogane, Y. Ando and T. Miyazaki, Applied Physics Letters **94** (12), 122503 (2009).
81. F. Wu, S. Mizukami, D. Watanabe, E. P. Sajitha, H. Naganuma, M. Oogane, Y. Ando and T. Miyazaki, IEEE Transactions on Magnetics **46** (6), 1863 (2010).
82. T. Kubota, Y. Miura, D. Watanabe, S. Mizukami, F. Wu, H. Naganuma, X. M. Zhang, M. Oogane, M. Shirai, Y. Ando and T. Miyazaki, Applied Physical Express **4** (4), 043002 (2011).
83. H. Kurt, K. Rode, M. Venkatesan, P. Stamenov and J. M. D. Coey, Physical Review B **83** (2), 020405 (2011).
84. K. Inomata, N. Tezuka, S. Okamura, H. Kurebayashi and A. Hirohata, Journal of Applied Physics **95** (11), 7234 (2004).
85. S. Okamura, R. Goto, S. Sugimoto, N. Tezuka and K. Inomata, Journal of Applied Physics **96** (11), 6561 (2004).
86. H. Kubota, J. Nakata, M. Oogane, Y. Ando, A. Sakuma and T. Miyazaki, Japanese Journal of Applied Physics **43**, L984-L986 (2004).
87. Y. Sakuraba, J. Nakata, M. Oogane, Y. Ando, H. Kato, A. Sakuma, T. Miyazaki and H. Kubota, Applied Physical Letters **88** (2), 022503 (2006).
88. S. Kammerer, A. Thomas, A. Hutten and G. Reiss, Applied Physical Letters **85** (1), 79 (2004).
89. Y. Sakuraba, J. Nakata, M. Oogane, H. Kubota, Y. Ando, A. Sakuma and T. Miyazaki, Japanese Journal of Applied Physics **44**, L1100-L1102 (2005).
90. Y. Sakuraba, M. Hattori, M. Oogane, Y. Ando, H. Kato, A. Sakuma, T. Miyazaki and H. Kubota, Applied Physical Letters **88** (19), 192508 (2006).
91. S. Okamura, A. Miyazaki, S. Sugimoto, N. Tezuka and K. Inomata, Applied Physical Letters **86** (23), 232503 (2005).
92. Z. Gercsi, A. Rajanikanth, Y. K. Takahashi, K. Hono, M. Kikuchi, N. Tezuka and K. Inomata, Applied Physical Letters **89** (8) (2006).
93. N. Tezuka, N. Ikeda, A. Miyazaki, S. Sugimoto, M. Kikuchi and K. Inomata, Applied Physical Letters **89** (11), 112514 (2006).
94. D. Ebke, J. Schmalhorst, N. N. Liu, A. Thomas, G. Reiss and A. Hutten, Applied Physics Letters **89** (16), 162506 (2006).
95. N. Tezuka, N. Ikeda, S. Sugimoto and K. Inomata, Applied Physics Letters **89** (25), 252508 (2006).
96. N. Tezuka, N. Ikeda, S. Sugimoto and K. Inomata, Japanese Journal of Applied Physics **46** (17-19), L454-L456 (2007).
97. H. Sukegawa, W. Wang, R. Shan, T. Nakatani, K. Inomata and K. Hono, Physical Review B **79** (18), 184418 (2009).
98. T. Marukame and M. Yamamoto, Journal of Applied Physics **101** (8), 083906 (2007).
99. T. Ishikawa, T. Marukame, H. Kijima, K.-I. Matsuda, T. Uemura, M. Arita and M. Yamamoto, Applied Physical Letters **89** (19), 192505 (2006).
100. T. Ishikawa, S. Hakamata, K.-i. Matsuda, T. Uemura and M. Yamamoto, Journal of Applied Physics **103** (7), 07A919 (2008).

101. S. Tsunegi, Y. Sakuraba, M. Oogane, K. Takanashi and Y. Ando, Applied Physical Letters **93** (11), 112506 (2008).
102. W. H. Butler, Science and Technology of Advanced Materials **9** (1), 014106 (2008).
103. Y. Miura, H. Uchida, Y. Oba, K. Abe and M. Shirai, Physical Review B **78** (6), 064416 (2008).
104. R. Matsumoto, A. Fukushima, T. Nagahama, Y. Suzuki, K. Ando and S. Yuasa, Applied Physics Letters **90** (25), 252506 (2007).
105. W. Wang, E. Liu, M. Kodzuka, H. Sukegawa, M. Wojcik, E. Jedryka, G. H. Wu, K. Inomata, S. Mitani and K. Hono, Physical Review B **81** (14), 140402 (2010).
106. M. Ležaić, P. Mavropoulos, J. Enkovaara, G. Bihlmayer and S. Blügel, Physical Review Letters **97** (2), 026404 (2006).
107. L. Chioncel, Y. Sakuraba, E. Arrigoni, M. I. Katsnelson, M. Oogane, Y. Ando, T. Miyazaki, E. Burzo and A. I. Lichtenstein, Physical Review Letters **100** (8), 086402 (2008).
108. K. Miyamoto, A. Kimura, Y. Miura, M. Shirai, M. Ye, Y. Cui, K. Shimada, H. Namatame, M. Taniguchi, Y. Takeda, Y. Saitoh, E. Ikenaga, S. Ueda, K. Kobayashi and T. Kanomata, Physical Review B **79** (10), 100405 (2009).
109. P. Mavropoulos, M. Ležaić and S. Blügel, Physical Review B **72** (17), 174428 (2005).
110. A. Sakuma, Y. Toga and H. Tsuchiura, Journal of Applied Physics **105** (7), 07C910 (2009).
111. Y. Miura, K. Abe and M. Shirai, Physical Review B **83** (21), 214411 (2011).
112. G. H. Fecher and C. Felser, Journal of Physics D: Applied Physics **40** (6), 1582 (2007).
113. T. M. Nakatani, A. Rajanikanth, Z. Gercsi, Y. K. Takahashi, K. Inomata and K. Hono, J Appl Phys **102** (3), 033916 (2007).
114. R. Shan, H. Sukegawa, W. H. Wang, M. Kodzuka, T. Furubayashi, T. Ohkubo, S. Mitani, K. Inomata and K. Hono, Physical Review Letters **102** (24), 246601 (2009).
115. N. Tezuka, N. Ikeda, F. Mitsuhashi and S. Sugimoto, Applied Physical Letters **94** (16), 162504 (2009).
116. W. H. Wang, H. Sukegawa and K. Inomata, Physical Review B **82** (9), 092402 (2010).
117. K. Nagasaka, Journal of Magnetism and Magnetic Materials **321** (6), 508 (2009).
118. Y. H. Chen, D. Song, J. M. Qiu, P. Kolbo, L. Wang, Q. He, M. Covington, S. Stokes, V. Sapozhnikov, D. Dimitrov, K. Z. Gao and B. Miller, IEEE Transactions on Magnetics **46** (3), 697 (2010).
119. M. Takagishi, K. Yamada, H. Iwasaki, H. N. Fuke and S. Hashimoto, IEEE Transactions on Magnetics **46** (6), 2086 (2010).
120. M. Covington, M. AlHajDarwish, Y. Ding, N. J. Gokemeijer and M. A. Seigler, Physical Review B **69** (18), 184406 (2004).
121. J. R. Childress, M. J. Carey, S. I. Kiselev, J. A. Katine, S. Maat and N. Smith, Journal of Applied Physics **99** (8), 08S305 (2006).
122. K. Yakushiji, K. Saito, S. Mitani, K. Takanashi, Y. K. Takahashi and K. Hono, Applied Physics Letters **88** (22), 222504 (2006).
123. Y. Sakuraba, T. Iwase, K. Saito, S. Mitani and K. Takanashi, Applied Physics Letters **94** (1), 012511 (2009).
124. V. Ko, G. Han, J. Qiu and Y. P. Feng, Applied Physics Letters **95** (20), 202502 (2009).
125. K. Nikolaev, P. Kolbo, T. Pokhil, X. Peng, Y. Chen, T. Ambrose and O. Mryasov, Applied Physics Letters **94** (22), 222501 (2009).
126. V. Ko, G. Han and Y. P. Feng, Journal of Magnetism and Magnetic Materials **322** (20), 2989

(2010).
127. Y. Sakuraba, K. Izumi, T. Iwase, S. Bosu, K. Saito, K. Takanashi, Y. Miura, K. Futatsukawa, K. Abe and M. Shirai, Physical Review B **82** (9), 094444 (2010).
128. Z. Q. Bai, Y. H. Lu, L. Shen, V. Ko, G. C. Han and Y. P. Feng, Journal of Applied Physics **111** (9), 093911 (2012).
129. T. Furubayashi, K. Kodama, H. Sukegawa, Y. K. Takahashi, K. Inomata and K. Hono, Applied Physics Letters **93** (12), 122507 (2008).
130. T. M. Nakatani, T. Furubayashi, S. Kasai, H. Sukegawa, Y. K. Takahashi, S. Mitani and K. Hono, Applied Physics Letters **96** (21), 212501 (2010).
131. T. M. Nakatani, S. Mitani, T. Furubayashi and K. Hono, Applied Physics Letters **99** (18), 182505 (2011).
132. T. Iwase, Y. Sakuraba, S. Bosu, K. Saito, S. Mitani and K. Takanashi, Appled Physics Express **2** (6), 063003 (2009).
133. Y. Sakuraba, K. Izumi, T. Iwase, S. Bosu, K. Saito, K. Takanashi, Y. Miura, K. Futatsukawa, K. Abe and M. Shirai, Physical Review B **82** (9), 094444 (2010).
134. N. Hase, T. M. Nakatani, S. Kasai, Y. K. Takahashi, T. Furubayashi and K. Hono, Journal of Magnetism and Magnetic Materials **324** (4), 440 (2012).
135. N. Hase, B. Varaprasad, T. M. Nakatani, H. Sukegawa, S. Kasai, Y. K. Takahashi, T. Furubayashi and K. Hono, Journal of Applied Physics **108** (9), 093916 (2010).
136. Y. K. Takahashi, A. Srinivasan, B. Varaprasad, A. Rajanikanth, N. Hase, T. M. Nakatani, S. Kasai, T. Furubayashi and K. Hono, Applied Physics Letters **98** (15), 152501 (2011).
137. J. Sato, M. Oogane, H. Naganuma and Y. Ando, Applied Physics Express **4** (11), 043002 (2011).
138. T. Mizuno, Y. Tsuchiya, T. Machita, S. Hara, D. Miyauchi, K. Shimazawa, T. Chou, K. Noguchi and K. Tagami, IEEE Transactions on Magnetics **44** (11), 3584 (2008).
139. K. Nikolaev, P. Anderson, P. Kolbo, D. Dimitrov, S. Xue, X. L. Peng, T. Pokhil, H. S. Cho and Y. Chen, Journal of Applied Physics **103** (7), 07F533 (2008).
140. K. Kodama, T. Furubayashi, H. Sukegawa, T. M. Nakatani, K. Inomata and K. Hono, Journal of Applied Physics **105** (7), 07E905 (2009).
141. T. Furubayashi, K. Kodama, T. M. Nakatani, H. Sukegawa, Y. K. Takahashi, K. Inomata and K. Hono, Journal of Applied Physics **107** (11), 113917 (2010).
142. M. J. Carey, S. Maat, S. Chandrashekariaih, J. A. Katine, W. Chen, B. York and J. R. Childress, Journal of Applied Physics **109** (9), 093912 (2011).
143. T. Ambrose and O. Mryasov, U.S. Patent No. 6876522 (5 April 2005).
144. V. Ko, J. Qiu, P. Luo, G. C. Han and Y. P. Feng, Journal of Applied Physics **109** (7), 07B103(2011).
145. Z. Bai, Y. Cai, L. Shen, G. Han and Y. Feng, in *ArXiv e-prints*, Vol. 1301, pp. 6106 (2013).
146. S. Chadov, T. Graf, K. Chadova, X. Dai, F. Casper, G. H. Fecher and C. Felser, Physical Review Letters **107** (4), 047202 (2011).
147. G. D. Fuchs, N. C. Emley, I. N. Krivorotov, P. M. Braganca, E. M. Ryan, S. I. Kiselev, J. C. Sankey, D. C. Ralph, R. A. Buhrman and J. A. Katine, Applied Physics Letters **85** (7), 1205 (2004).
148. Y. Huai, F. Albert, P. Nguyen, M. Pakala and T. Valet, Applied Physics Letters **84** (16), 3118 (2004).
149. Z. Diao, D. Apalkov, M. Pakala, Y. Ding, A. Panchula and Y. Huai, Applied Physics Letters **87** (23), 232502 (2005).

150. J. Hayakawa, S. Ikeda, Y. M. Lee, R. Sasaki, T. Meguro, F. Matsukura, H. Takahashi and H. Ohno, Japanese Journal of Applied Physics **44** (37-41), L1267-L1270 (2005).
151. H. Kubota, A. Fukushima, Y. Ootani, S. Yuasa, K. Ando, H. Maehara, K. Tsunekawa, D. D. Djayaprawira, N. Watanabe and Y. Suzuki, Japanese Journal of Applied Physics **44** (37-41), L1237-L1240 (2005).
152. M. Pakala, Y. Huai, T. Valet, Y. Ding and Z. Diao, Journal of Applied Physics **98** (5), 056107 (2005).
153. S. Ikeda, K. Miura, H. Yamamoto, K. Mizunuma, H. D. Gan, M. Endo, S. Kanai, J. Hayakawa, F. Matsukura and H. Ohno, Nature Materials **9** (9), 721 (2010).
154. R. Sbiaa, H. Meng and S. N. Piramanayagam, Physica Status Solidi-Rapid Research Letters **5** (12), 413 (2011).
155. N. Nishimura, T. Hirai, A. Koganei, T. Ikeda, K. Okano, Y. Sekiguchi and Y. Osada, Journal of Applied Physics **91** (8), 5246 (2002).
156. O. Redon, B. Dieny and B. Rodmacq, U.S. Patent No. 6532164 B2 (3 November 2003).
157. A. D. Kent, B. Ozyilmaz and E. del Barco, Applied Physics Letters **84** (19), 3897 (2004).
158. K. J. Lee, O. Redon and B. Dieny, Applied Physics Letters **86** (2), 022505 (2005).
159. L. Liu, T. Moriyama, D. C. Ralph and R. A. Buhrman, Applied Physics Letters **94** (12), 122508 (2009).
160. T. Moriyama, T. J. Gudmundsen, P. Y. Huang, L. Liu, D. A. Muller, D. C. Ralph and R. A. Buhrman, Applied Physics Letters **97** (7), 072513 (2010).
161. P. K. Amiri, Z. M. Zeng, J. Langer, H. Zhao, G. Rowlands, Y. J. Chen, I. N. Krivorotov, J. P. Wang, H. W. Jiang, J. A. Katine, Y. Huai, K. Galatsis and K. L. Wang, Applied Physics Letters **98** (11), 112507 (2011).
162. M. T. Rahman, A. Lyle, P. K. Amiri, J. Harms, B. Glass, H. Zhao, G. Rowlands, J. A. Katine, J. Langer, I. N. Krivorotov, K. L. Wang and J. P. Wang, Journal of Applied Physics **111** (7), 07C907 (2012).
163. S. Mangin, D. Ravelosona, J. A. Katine, M. J. Carey, B. D. Terris and E. E. Fullerton, Nat Mater **5** (3), 210 (2006).
164. K. Mizunuma, S. Ikeda, J. H. Park, H. Yamamoto, H. Gan, K. Miura, H. Hasegawa, J. Hayakawa, F. Matsukura and H. Ohno, Applied Physics Letters **95** (23), 232516 (2009).
165. K. Yakushiji, T. Saruya, H. Kubota, A. Fukushima, T. Nagahama, S. Yuasa and K. Ando, Applied Physics Letters **97** (23), 232508 (2010).
166. M. Yoshikawa, E. Kitagawa, T. Nagase, T. Daibou, M. Nagamine, K. Nishiyama, T. Kishi and H. Yoda, IEEE Transactions on Magnetics **44** (11), 2573 (2008).
167. G. Kim, Y. Sakuraba, M. Oogane, Y. Ando and T. Miyazaki, Applied Physics Letters **92** (17), 172502 (2008).
168. H. Ohmori, T. Hatori and S. Nakagawa, Journal of Applied Physics **103** (7), 07A911-913 (2008).
169. S. Mizukami, S. Iihama, N. Inami, T. Hiratsuka, G. Kim, H. Naganuma, M. Oogane and Y. Ando, Applied Physics Letters **98** (5), 052501 (2011).
170. B. Rodmacq, A. Manchon, C. Ducruet, S. Auffret and B. Dieny, Physical Review B **79** (2), 024423 (2009).
171. J. H. Jung, S. H. Lim and S. R. Lee, Applied Physics Letters **96** (4), 042503 (2010).
172. H. Sukegawa, Z. Wen, K. Kondou, S. Kasai, S. Mitani and K. Inomata, Appl Phys Lett **100** (18), 182403 (2012).

173. Z. C. Wen, H. Sukegawa, S. Kasai, M. Hayashi, S. Mitani and K. Inomata, Applied Physics Express **5** (6), 063003 (2012).
174. S. Mizukami, F. Wu, A. Sakuma, J. Walowski, D. Watanabe, T. Kubota, X. Zhang, H. Naganuma, M. Oogane, Y. Ando and T. Miyazaki, Physical Review Letters **106** (11), 117201 (2011).
175. Q. L. Ma, T. Kubota, S. Mizukami, X. M. Zhang, H. Naganuma, M. Oogane, Y. Ando and T. Miyazaki, Applied Physics Letters **101** (3), 032402 (2012).
176. T. Kubota, M. Araidai, S. Mizukami, X. Zhang, Q. Ma, H. Naganuma, M. Oogane, Y. Ando, M. Tsukada and T. Miyazaki, Applied Physics Letters **99** (19), 192509 (2011).
177. C. Yongqing, B. Zhaoqiang, Y. Ming and F. Yuan Ping, EPL (Europhysics Letters) **99** (3), 37001 (2012).
178. P. Klaer, C. A. Jenkins, V. Alijani, J. Winterlik, B. Balke, C. Felser and H. J. Elmers, Applied Physics Letters **98** (21), 212510 (2011).
179. V. Alijani, J. Winterlik, G. H. Fecher and C. Felser, Applied Physics Letters **99** (22), 222510 (2011).
180. A. K. Nayak, C. Shekhar, J. Winterlik, A. Gupta and C. Felser, Applied Physics Letters **100** (15), 152404 (2012).
181. J. C. Slater, Physical Review **49** (12), 931 (1936).
182. L. Pauling, Physical Review **54** (11), 899 (1938).
183. J. Winterlik, S. Chadov, A. Gupta, V. Alijani, T. Gasi, K. Filsinger, B. Balke, G. H. Fecher, C. A. Jenkins, F. Casper, J. Kübler, G.-D. Liu, L. Gao, S. S. P. Parkin and C. Felser, Advanced Materials **24** (47), 6283 (2012).
184. J. C. Suits, Solid State Communications **18** (3), 423 (1976).
185. L. Van Hove, Physical Review **89** (6), 1189 (1953).
186. N. P. Butch, P. Syers, K. Kirshenbaum, A. P. Hope and J. Paglione, Physical Review B **84** (22), 220504(R) (2011).
187. K. Gofryk, D. Kaczorowski, T. Plackowski, A. Leithe-Jasper and Y. Grin, Physical Review B **84** (3), 035208 (2011).
188. A. Kreyssig, M. G. Kim, J. W. Kim, D. K. Pratt, S. M. Sauerbrei, S. D. March, G. R. Tesdall, S. L. Bud'ko, P. C. Canfield, R. J. McQueeney and A. I. Goldman, Physical Review B **84** (22), 220408(R) (2011).
189. C. Li, J. S. Lian and Q. Jiang, Physical Review B **83** (23), 235125 (2011).
190. C. Liu, Y. Lee, T. Kondo, E. D. Mun, M. Caudle, B. N. Harmon, S. L. Bud'ko, P. C. Canfield and A. Kaminski, Physical Review B **83** (20), 205133 (2011).
191. S. Ouardi, C. Shekhar, G. H. Fecher, X. Kozina, G. Stryganyuk, C. Felser, S. Ueda and K. Kobayashi, Applied Physics Letters **98** (21), 211901 (2011).
192. H. J. Zhang, S. Chadov, L. Muchler, B. H. Yan, X. L. Qi, J. Kubler, S. C. Zhang and C. Felser, Physical Review Letters **106** (15), 156402 (2011).
193. X. M. Zhang, W. H. Wang, E. K. Liu, G. D. Liu, Z. Y. Liu and G. H. Wu, Applied Physics Letters **99** (7), 071901 (2011).
194. F. Casper, T. Graf, S. Chadov, B. Balke and C. Felser, Semiconductor Science and Technology **27** (6), 063001 (2012).
195. T. Miyawaki, N. Sugimoto, N. Fukatani, T. Yoshihara, K. Ueda, N. Tanaka and H. Asano, Journal of Applied Physics **111** (7), 07E327 (2012).
196. L. Muechler, H. J. Zhang, S. Chadov, B. H. Yan, F. Casper, J. Kubler, S. C. Zhang and C. Felser,

Angewandte Chemie-International Edition **51** (29), 7221 (2012).
197. X. M. Zhang, G. D. Liu, Y. Du, E. K. Liu, W. H. Wang, G. H. Wu and Z. Y. Liu, Acta Physica Sinica **61** (12), 123101 (2012).
198. M. Z. Hasan and C. L. Kane, Reviews of Modern Physics **82** (4), 3045 (2010).
199. X.-L. Qi and S.-C. Zhang, Reviews of Modern Physics **83** (4), 1057 (2011).
200. H. Steinberg, D. R. Gardner, Y. S. Lee and P. Jarillo-Herrero, Nano Letters **10** (12), 5032 (2010).
201. O. V. Yazyev, J. E. Moore and S. G. Louie, Physical Review Letters **105** (26), 266802 (2010).
202. Y. H. Lu, unpublished (2012).
203. A. Roy, J. W. Bennett, K. M. Rabe and D. Vanderbilt, Physical Review Letters **109** (3), 037602 (2012).
204. B. Sa, J. Zhou, Z. Sun, J. Tominaga and R. Ahuja, Physical Review Letters **109** (9), 096802 (2012).

**Figures**

Fig. 1. Periodic table of Heusler compounds. The huge number of full Heusler compounds can be formed by combination of the different elements according to the color scheme. The electronegativity value is given below the element symbol. For half Heusler compounds XYZ, the elements are ordered according to their electronegtivity.

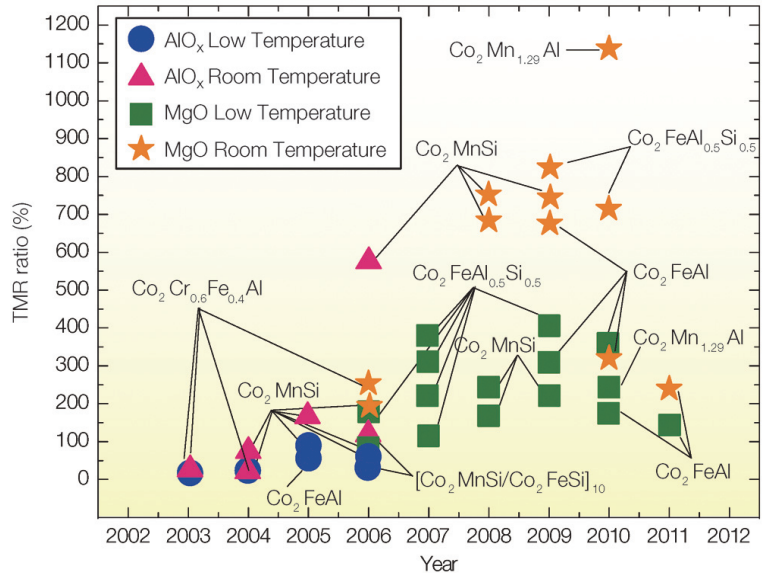

Fig. 2. Development of the TMR ratio for MTJs with Heusler electrodes at low temperatures (blue symbols) and room temperature (red symbols). Data taken from Ref. 68 and Refs. 84-105.

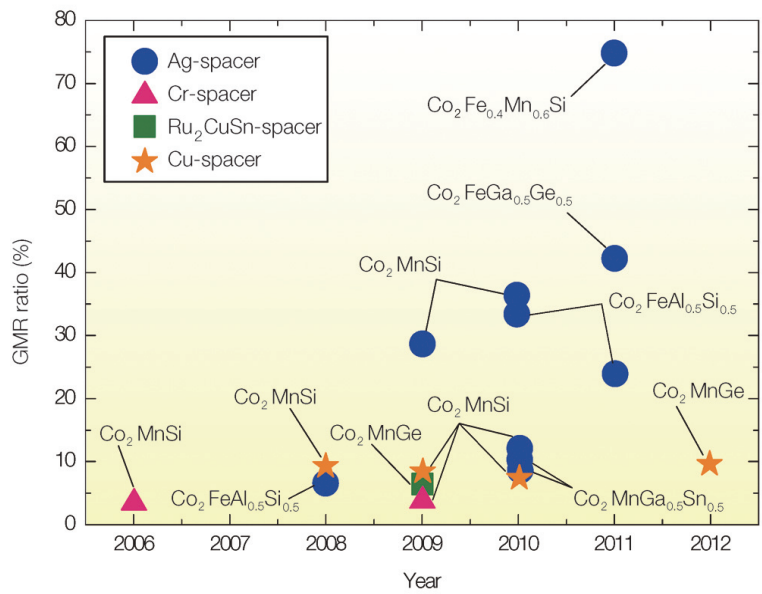

Fig. 3. Development of the GMR ratio at room temperature for CPP-GMR SVs with Heusler electrodes. Data taken from Refs. 122-141.

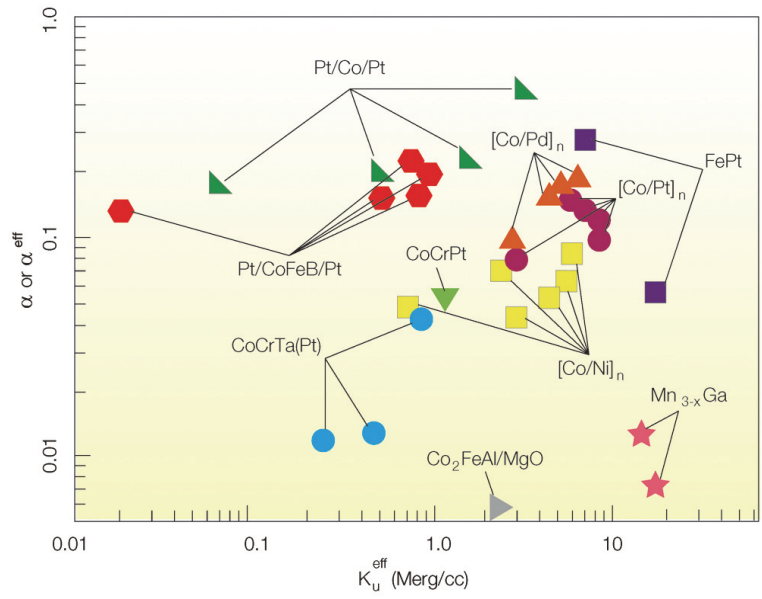

Fig. 4. A collection of Gilbert damping constants and uniaxial magnetic anisotropy constants in some PMA films.

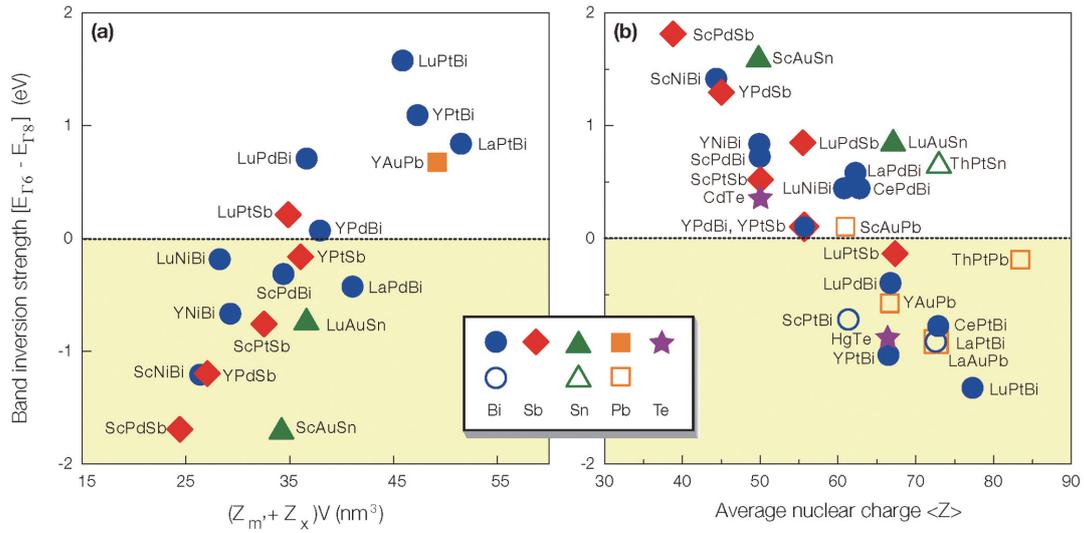

Fig. 5 Topological half-Heusler family of compounds. Topological band inversion strengths are plotted as a function of the product of the sum of the nuclear charges for several classes of half-Heuslers. The inversion strength is defined as the energy difference between the $\Gamma_8$ and $\Gamma_6$ states at the $\Gamma$ point. Negative values denote absence of inversion. Materials with positive band inversion strength are candidates for topological insulators once the lattice symmetry is broken. Data of (a) taken from Ref. 53; Data of (b) taken from Ref. 51.

Bai_Figure1

# PERIODIC TABLE OF HEUSLER COMPOUNDS

| 1 H 2.20 | | ATOMIC NUMBER 29 Cu 1.90 — SYMBOL ELECTRONEGATIVITY | | | $X_2YZ$ Full Heusler* XYZ Half Heusler# | | | | | | | | | | | | 2 He |
|---|---|---|---|---|---|---|---|---|---|---|---|---|---|---|---|---|---|
| 3 Li 0.98 | 4 Be 1.57 | | | | | | | | | | 5 B 2.04 | 6 C 2.55 | 7 N 3.04 | 8 O 3.44 | 9 F 3.98 | | 10 Ne |
| 11 Na 0.93 | 12 Mg 1.31 | | | | | | | | | | 13 Al 1.61 | 14 Si 1.90 | 15 P 2.19 | 16 S 2.58 | 17 Cl 3.16 | | 18 Ar |
| 19 K 0.82 | 20 Ca 1.00 | 21 Sc 1.36 | 22 Ti 1.54 | 23 V 1.63 | 24 Cr 1.66 | 25 Mn 1.55 | 26 Fe 1.83 | 27 Co 1.88 | 28 Ni 1.91 | 29 Cu 1.90 | 30 Zn 1.65 | 31 Ga 1.81 | 32 Ge 2.01 | 33 As 2.18 | 34 Se 2.55 | 35 Br 2.96 | 36 Kr 3.00 |
| 37 Rb 0.82 | 38 Sr 0.95 | 39 Y 1.22 | 40 Zr 1.33 | 41 Nb 1.60 | 42 Mo 2.16 | 43 Te 1.90 | 44 Ru 2.20 | 45 Rh 2.28 | 46 Pd 2.20 | 47 Ag 1.93 | 48 Cd 1.69 | 49 In 1.78 | 50 Sn 1.96 | 51 Sb 2.05 | 52 Te 2.10 | 53 I 2.66 | 54 Xe 2.60 |
| 55 Cs 0.79 | 56 Ba 0.89 | 57-71 LANTHANIDE | 72 Hf 1.30 | 73 Ta 1.50 | 74 W 1.70 | 75 Re 1.90 | 76 Os 2.20 | 77 Ir 2.20 | 78 Pt 2.20 | 79 Au 2.40 | 80 Hg 1.90 | 81 Tl 1.80 | 82 Pb 1.80 | 83 Bi 1.90 | 84 Po 2.00 | 85 At 2.20 | 86 Rn |
| 87 Fr 0.70 | 88 Ra 0.90 | 89-103 ACTINIDE | | | | | | | | | | | | | | | |

LANTHANIDE

| 57 La 1.10 | 58 Ce 1.12 | 59 Pr 1.13 | 60 Nd 1.14 | 61 Pm 1.13 | 62 Sm 1.17 | 63 Eu 1.20 | 64 Gd 1.20 | 65 Tb 1.10 | 66 Dy 1.22 | 67 Ho 1.23 | 68 Er 1.24 | 69 Tm 1.25 | 70 Yb 1.10 | 71 Lu 1.27 |
|---|---|---|---|---|---|---|---|---|---|---|---|---|---|---|

ACTINIDE

| 89 Ac 1.10 | 90 Th 1.30 | 91 Pa 1.50 | 92 U 1.70 | 93 Np 1.30 | 94 Pu 1.28 | 95 Am 1.13 | 96 Cm 1.28 | 97 Bk 1.30 | 98 Cf 1.30 | 99 Es 1.30 | 100 Fm 1.30 | 101 Md 1.30 | 102 No 1.30 | 103 Lr 1.30 |
|---|---|---|---|---|---|---|---|---|---|---|---|---|---|---|

IUPAC nomenclature^ of Heusler compounds:

* Normally, the element (exists twice) is put at the beginning of the formula, whereas the main group element is placed at the end, such as $Co_2FeSi$.
  Exceptionally, the element (existes once) is put at the beginning of the formula if such element can definitively be defined to be most electropositive, for instance $LiCu_2Sb$.

# The elements are ordered according to their electronegtivity. The most electropositive element is put at the beginning of the formula, for example MnNiSb.

^ http://www.chem.qmul.ac.uk/iupac/

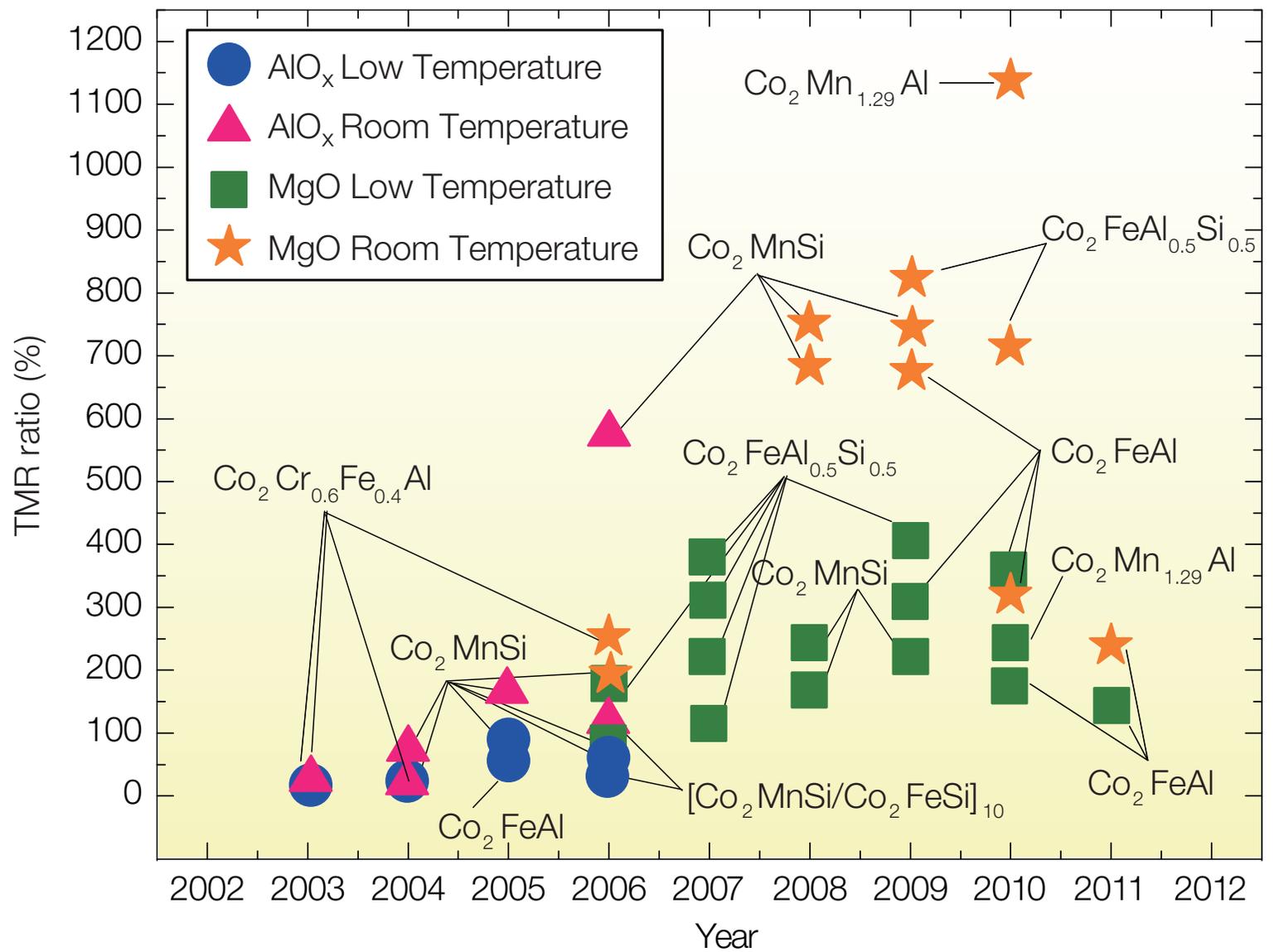

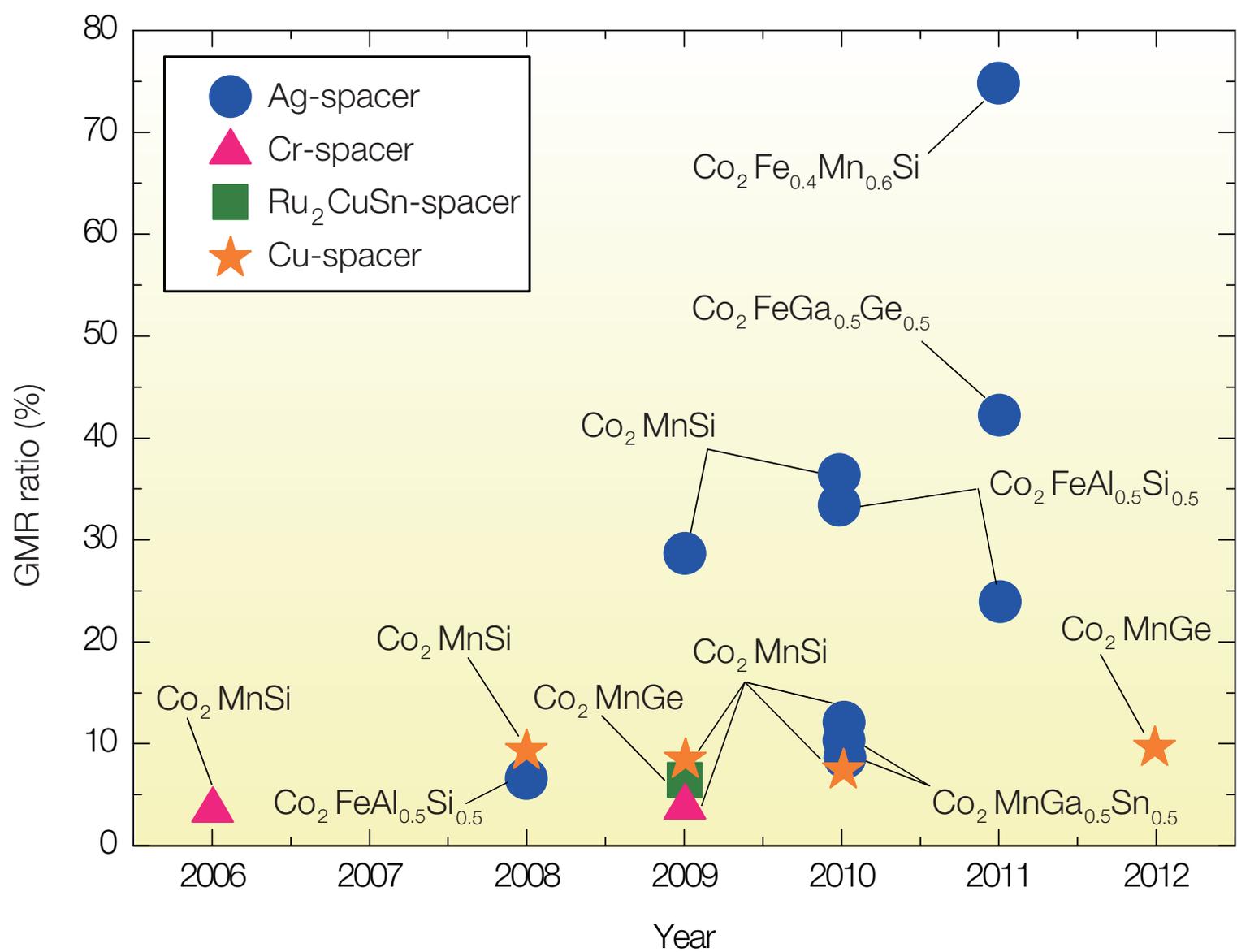

Bai_Figure4

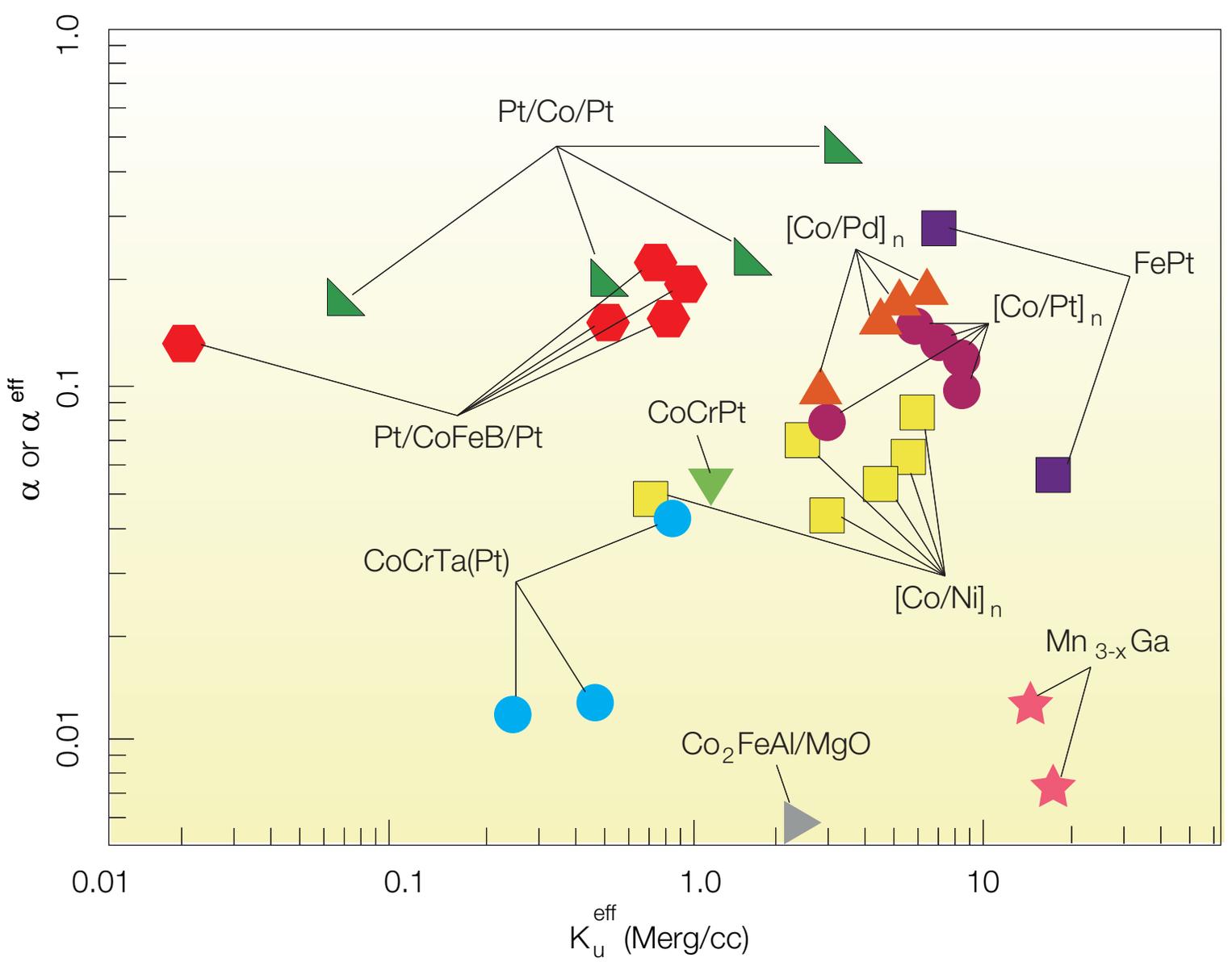

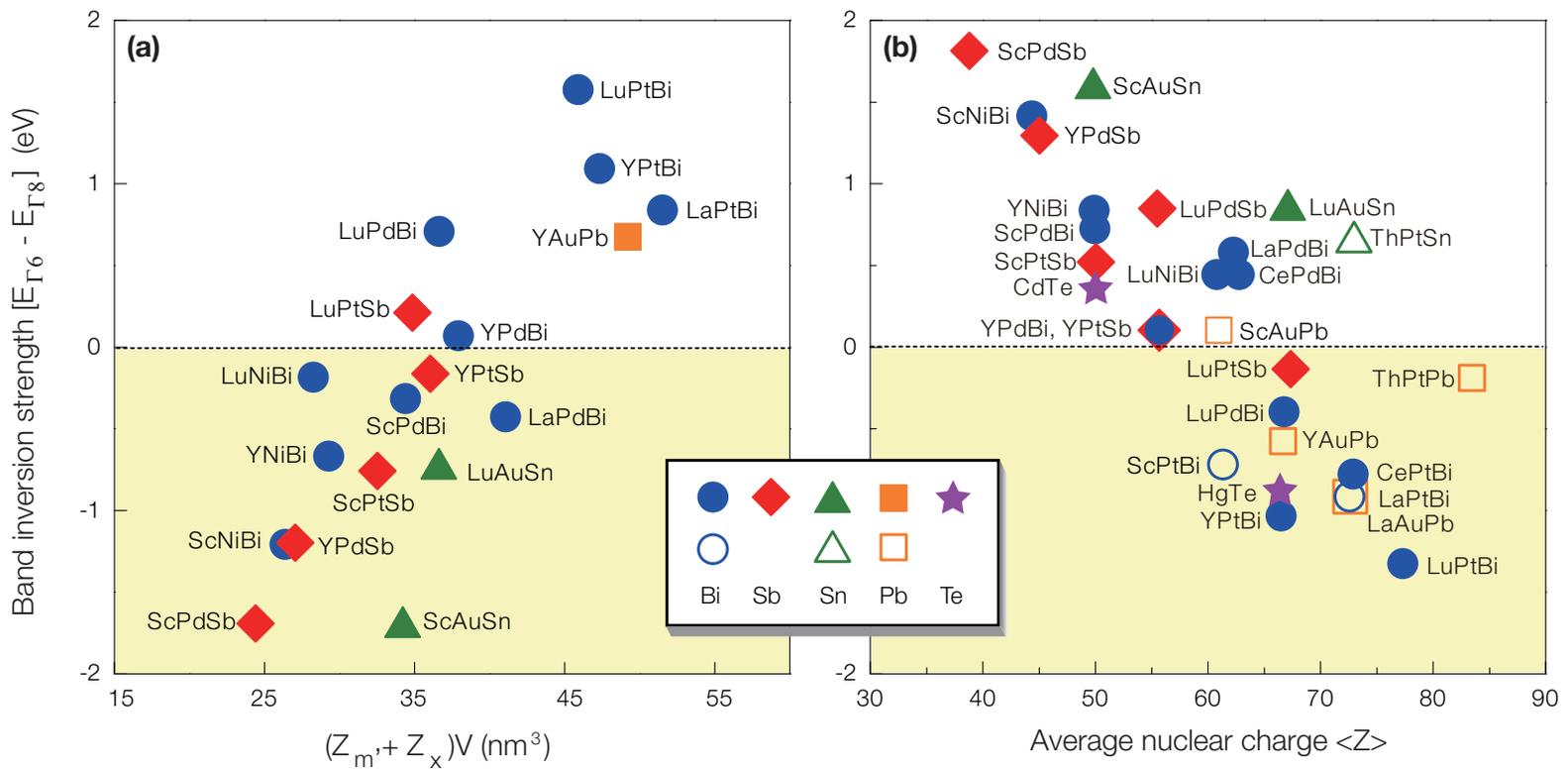